\documentclass[aps,preprint,showpacs,floatfix]{revtex4}
\usepackage{graphicx,bm,amsmath,dcolumn}
\begin{document}
\newcommand{\half}{\frac{1}{2}}
\newcommand{\ith}{^{(i)}}
\newcommand{\im}{^{(i-1)}}
\newcommand{\gae}
{\,\hbox{\lower0.5ex\hbox{$\sim$}\llap{\raise0.5ex\hbox{$>$}}}\,}
\newcommand{\lae}
{\,\hbox{\lower0.5ex\hbox{$\sim$}\llap{\raise0.5ex\hbox{$<$}}}\,}

\title{Critical properties of a dilute O($n$) model on the kagome lattice}
\author{Biao Li~$^{1}$,
Wenan Guo~$^{1}$ and Henk W.J. Bl\"ote~$^{2,3}$} 
\affiliation{$^{1}$Physics Department, Beijing Normal University,
Beijing 100875, P. R. China }
\affiliation{$^{2}$Faculty of Applied Sciences, Delft University of
Technology, P.O. Box 5046, 2600 GA Delft, The Netherlands}
\affiliation{$^{3}$ Instituut Lorentz, Leiden University,
  P.O. Box 9506, 2300 RA Leiden, The Netherlands}

\date{\today} 
\begin{abstract}
A critical dilute O($n$) model on the kagome lattice is investigated
analytically and numerically. 
We employ a number of exact equivalences which, in a few steps, link the
critical O($n$) spin model on the kagome lattice to the exactly solvable 
critical $q$-state Potts model on the honeycomb lattice with $q=(n+1)^2$.
The intermediate steps involve the random-cluster model on the honeycomb
lattice, and a fully packed loop model with loop weight $n'=\sqrt{q}$
and a dilute loop model with loop weight $n$, both on the kagome lattice.
This mapping enables the determination of a branch of critical points 
of the dilute O($n$) model, as well as some of its critical properties.
For $n=0$, this model reproduces the known universal properties of the
$\theta$ point describing the collapse of a polymer.  
For $n\neq 0$ it displays a line of multicritical points, with 
the same universal properties as a branch of critical behavior that was
found earlier in a dilute O($n$) model on the square lattice. 
These findings are supported by a finite-size-scaling
analysis in combination with transfer-matrix calculations.  
\end{abstract}
\maketitle 
\section{Introduction}
\label{intro}
The first exact results \cite{N} for the O($n$) critical properties
were obtained for a model on the honeycomb lattice, and revealed not only
the critical point, but also some universal parameters of the critical
state, as well as the low-temperature phase, as a function of $n$.
The derivation of these results depends on a special choice of the
O($n$)-symmetric interaction between the $n$-component spins of the
O($n$) model, which enables a mapping on a loop gas \cite{DMNS}.
These results were supposed to apply to a whole universality class of 
O($n$)-symmetric models in two dimensions.

Since then, also O($n$) models on the square and triangular lattices
were investigated \cite{BN,KBN}. Indeed, branches were found with the 
same universal properties as the honeycomb model, but in addition to
these, several other branches of critical behavior were reported.
Among these, we focus on `branch 0' as reported in Refs.~\cite{BN,KBN}.
The points on this branch appear to describe a higher critical point.
For $n=0$, it can be identified with the so-called $\theta$ point \cite{DS}
describing the collapse of a polymer in two dimensions, which has been
interpreted as a tricritical O($n=0$) model. It has indeed been found
that the introduction of a sufficiently strong and suitably chosen 
attractive potential between the loop segments changes the ordinary 
O($n=0$) transition into a first-order one \cite{BBN}, such that this
change precisely coincides with the $n=0$ point of branch 0. Thus, the
$\theta$ point plays the role of a tricritical O($n=0$) transition.
Furthermore, it has been verified that tricriticality in the O($n$)
model can be introduced by adding a sufficient concentration of
vacancies into the system \cite{GNB}. More precisely, the
introduction of vacancies leads to a branch of higher critical points,
of which the points $n=0$  and $n=1$ belong to universality  classes
(of the $\theta$ point and the tricritical Blume-Capel model respectively)
that have been described earlier as tricritical points.

However, the critical points of branch 0 on the square lattice appear to
display universal properties that are different from those of the branch
of higher critical points of the O($n$) model with vacancies \cite{GNB},
except at the intersection point of the two branches at $n=0$.
It thus appears that the continuation of the $\theta$ point at $n=0$ to
$n\ne 0$ can be done in different ways, leading to different universality
classes. In order to gain further insight in this situation, the present
work considers an O($n$) loop model on the kagome lattice with the
purpose to find a $\theta$-like point, to continue this point to $n\ne 0$
and to explore the resulting universality.

\section{Mappings}
\label{mapsec} 
The partition function of the spin representation of $q$-state Potts
model on the honeycomb lattice
\begin{equation}
Z_{\rm Potts}=\sum_{\{S\}}\exp \left(K\sum_{<i,j>}\delta_{s_i,s_j}\right)
\label{spinpotts}
\end{equation}
depends on the temperature $T$ by the coupling $K=J/k_{\rm B}T$,
where $J$ is the nearest-neighbor spin-spin interaction. The spins
$s_i$ can assume values 1, 2, $\cdots$, $q$ and their index $i$ 
labels the sites of the honeycomb lattice.
The first summation is over all possible spin configurations $\{S\}$,
and the second one is over the nearest neighbor spin pairs.
This Potts model can be subjected to a series of mappings which 
lead, via the  random-cluster model and a fully-packed loop model, 
to a dilute O($n$) loop model which can also be interpreted as an
O($n$) spin model.

\subsection{Honeycomb Potts model to fully-packed kagome loop model}
\begin{figure}
\includegraphics[scale=0.3]{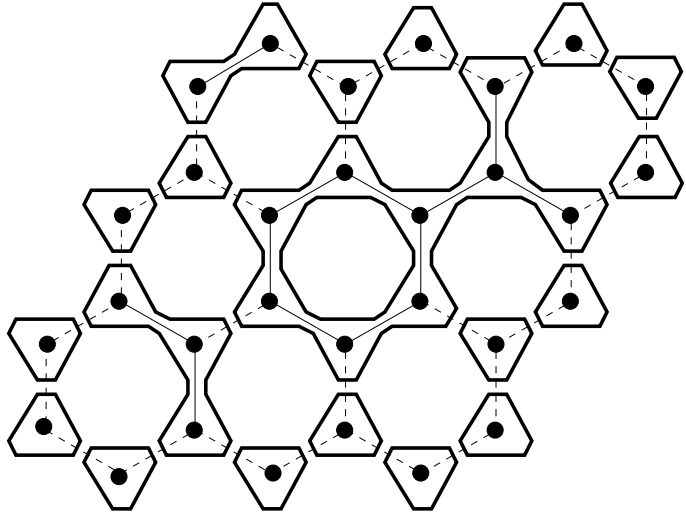}
\caption{Mapping of the RC model onto a FPL model.
The sites of the honeycomb lattice are shown as black circles.
The dashed and the thin solid lines display the empty and the occupied 
edges (bonds) of the RC model on the honeycomb lattice respectively.
The RC configuration is here represented by an FPL configuration on the
surrounding lattice, i.e., the kagome lattice. Its loops (bold solid
lines) follow the boundaries of the random clusters, both externally
and internally.
The Boltzmann weight of this finite-size configuration of the RC 
configuration is $u^{12}q^{19}$ according to Eq.~(\ref {zrc}), and that
of the corresponding FPL configuration is $a_1^{12}a_2^{26}n^{20}$
according to Eq.~(\ref {zfpl}).
}
\label{mapping}
\end{figure}

The introduction of bond variables, and a summation on the spin
variables map the Potts model onto the random-cluster (RC)
model \cite{KF}, with partition function
\begin{equation}
Z_{\rm RC}(u,q)=\sum_{\mathcal{B}}u^{N_b}q^{N_c} \, ,
\label{zrc}
\end{equation}
where $N_b$ is the number of bonds, $N_c$ the number of clusters, 
and $u \equiv e^K-1$ the weight of a bond. The sum is on all
configurations ${\mathcal{B}}$ of bond variables: each bond variable 
is either 1 (present) or 0 (absent). In Eq.~(\ref{zrc}),
$q$ can be considered a continuous real number, playing the
role of the weight of a cluster. Here, a cluster is either a single
site or a group of sites connected together by bonds on the lattice. 
A typical configuration of the RC model on the honeycomb
lattice is shown in Fig.~\ref{mapping}.

The next step is a mapping of the RC model on the honeycomb lattice
onto a fully packed loop (FPL) model on the kagome lattice, which
proceeds similarly as in the case of the square lattice \cite{BKW}.
The sites of FPL model sit in the middle of the edges of the 
honeycomb lattice, and thus form a kagome lattice \cite{IS}.
Fully packed here means that all edges of the kagome lattice
are covered by loop segments. The one-to-one correspondence between
these two configurations is established by requiring that the loops
do not intersect the occupied edges (bonds) of
the honeycomb RC model, and always intersect the empty edges, as  
illustrated in Fig.~\ref{mapping}. 

\begin{figure}
\includegraphics[scale=0.5]{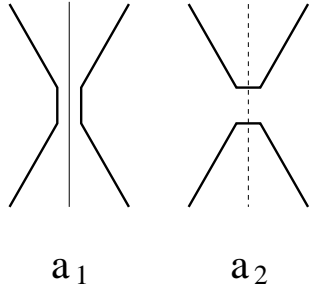}
\caption{Vertex weights of the FPL model. The bold solid lines represent
loop segments. The weight of vertex where the loops do not intersect a
bond (thin solid line) is $a_1$. The weight of a vertex where two loops
intersect an unoccupied edge (dashed line) is $a_2$. }
\label{2vs}
\end{figure}

To specify the Boltzmann weights of the FPL model, we assign a weight $n$
to each loop, a weight $a_1$ to each vertex where the loop segments do
not intersect an edge which is occupied by a bond of the RC model, 
and a weight $a_2$ to each vertex where the loop segments intersect an
edge which is empty in the RC model, as illustrated in Fig.~\ref{2vs}.
The partition function of the FPL model on the kagome lattice is thus
defined as
\begin{equation}
Z_{\rm FPL}^{\rm kag}(a_1, a_2, n)=
      \sum_{\mathcal{F}} a_1^{m_1}a_2^{m_2}n^{m_l} \, ,
\label{zfpl}
\end{equation}
where $m_1$ is the number of type-1 vertices, $m_2$ is the number of
type-2 vertices and $m_l$ the number of loops.
The sum is on all configurations $\mathcal{F}$ of loops covering all
the edges of the kagome lattice.

The one-to-one correspondence between RC configurations and FPL
configurations makes it possible to express the configuration parameters
$m_1$, $m_2$ and $m_l$ of the FPL in those of the RC model, namely
$N_b$ and $N_c$.
Each vertex of type-1 corresponds with a bond of the RC model on the
honeycomb lattice, thus
\begin{equation} 
m_1=N_b \, .
\label{e1}
\end{equation}
The total number of the two kinds of vertices is equal to
the number of edges on the honeycomb lattice, i.e.,
\begin{equation} 
m_1+m_2=\frac{3N}{2} \, ,
\label{e2}
\end{equation}
where $N$ is the total number of sites of the honeycomb lattice. 
Here we ignore surface effects of finite lattices.
Furthermore, a loop on the kagome lattice is either one surrounding a
random cluster on the honeycomb lattice, or one following the inside
of a loop formed by the bonds of a random cluster. Thus
\begin{equation} 
m_l=N_c+N_l \, ,
\label{e3}
\end{equation}
where $N_l$ is the loop number of the RC model. 
Together with the Euler relation
\begin{equation} 
N_c=N-N_b+N_l \, ,
\label{e4}
\end{equation}
Eqs.~(\ref{e1}) to (\ref{e3}) yield the numbers of vertices and loops 
on the kagome lattice as
\begin{eqnarray}
 m_1 &=& N_b \nonumber\\
 m_2 &=& 3N/2-N_b \label{rcfpl}\\
 m_l &=& 2N_c+N_b-N \, .\nonumber
\end{eqnarray}
Substitution in the partition function (\ref{zfpl}) leads to
\begin{equation}
Z_{\rm FPL}^{\rm kag}=
\left(\frac{a_2^{\frac{3}{2}}}{n}\right)^N\sum_{\mathcal{F}}
\left(\frac{a_1n}{a_2}\right)^{N_b}(n^2)^{N_c} \, .
\label{zfpl1}
\end{equation}
The weight of a given loop configuration is thus equal to the
corresponding RC weight $u^{N_b}q^{N_c}$ if
\begin{eqnarray}
n    &=& \sqrt{q} \nonumber\\
a_1  &=& u q^{-\frac{1}{6}} \label{wfpl}\\
a_2  &=& q^{\frac{1}{3}}  \, , \nonumber
\end{eqnarray}
which completes the mapping of the RC onto the FPL model.

\subsection{Fully-packed loop model to dilute loop model}
\label{fptodl}
\begin{figure}
\includegraphics[scale=0.45]{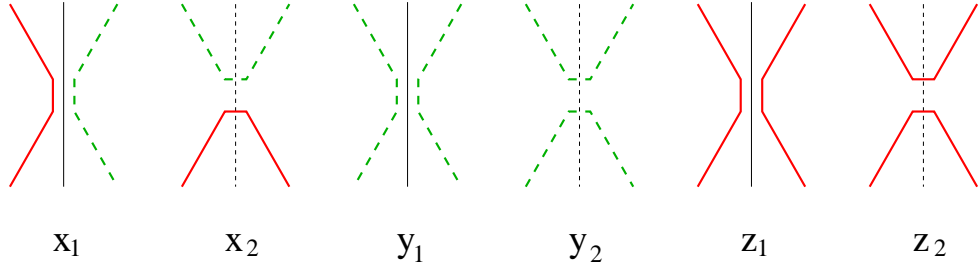}
\caption{(color online). Weights of colored vertices. The vertical solid
lines represent occupied edges (bonds) on the honeycomb lattice, while
broken lines stand for empty edges. The bold solid lines represent the
red loop segments, and the bold dashed lines the green ones.}
\label{6vs}
\end{figure}

\begin{figure}
\includegraphics[scale=0.45]{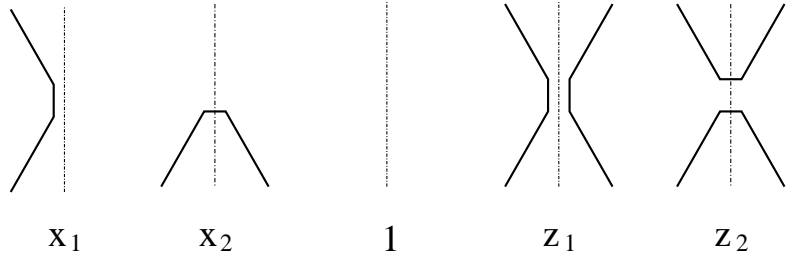}
\caption{The five vertex weights for the dilute loop model. The vertex
with weight 1 results from a summation involving the weights of vertices
3 and 4 in Fig.~\ref{6vs}.} 
\label{5vsdl}
\end{figure}

Next we map the FPL model on the kagome lattice onto a dilute loop
(DL) model on the same lattice, using a method due to Nienhuis 
(see e.g. Ref.~\cite{BN}). 
The partition function of the FPL model on the kagome lattice
is slightly rewritten as 
\begin{equation}
Z_{\rm FPL}^{\rm kag}=(a_1+a_2)^{\frac{3N}{2}}
\sum_{{\mathcal{F}}}w_1^{m_1}w_2^{m_2}[(n-1)+1]^{m_l}
\label{zfpl2}
\end{equation}
with $w_1=a_1 \big/ (a_1+a_2)$ and $w_2=a_2 \big/ (a_1+a_2)$.
Eq.~(\ref{zfpl2}) invites an interpretation  in terms of colored
loops, say red with loops of weight $n-1$ and green loops of weight 1. 
Each of the $2^{m_l}$ terms in the expansion of $[(n-1)+1]^{m_l}$ thus
specifies a way to color the loops:
$$[(n-1)+1]^{m_l}=\sum _{\{\rm colorings\}}(n-1)^{l_r}1^{l_g} \, ,$$ 
where $l_r$ and $l_g$ denote the number of red loops 
and green loops respectively, $l_r+l_g=m_l$.
Let $\mathcal{C}$ denote a graph $\mathcal{F}$ in which the colors of
all loops are specified. The partition sum can thus be expressed in terms 
of a summation over all colored loop configurations $\mathcal{C}$. 
The vertices of the kagome lattice are visited by two colored loops,
and can thus be divided into 6 types, shown in Fig.~\ref {6vs} with 
their associated weights $x_1=y_1=z_1=w_1$ and $x_2=y_2=z_2=w_2$. 
Thus, Eq.~(\ref {zfpl2}) assumes the form
\begin{equation}
Z^{\rm kag}_{\rm FPL}=(a_1+a_2)^{\frac{3N}{2}}
\sum_{\mathcal{C}}x_1^{N_{x_1}}x_2^{N_{x_2}}y_1^{N_{y_1}}y_2^{N_{y_2}}
z_1^{N_{z_1}}z_2^{N_{z_2}}(n-1)^{l_r}1^{l_g} \, .
\end{equation}
The sum $\sum_{\mathcal{C}}$ on all colored loop configurations may now be
replaced by two nested sums, the first of which is a sum
$\sum_{\mathcal{R}}$ on all dilute loop configurations 
of red loops, and the second sum
$\sum_{\mathcal{G}|\mathcal{R}}$ 
is on all configurations of green loops $\mathcal{G}$ that are 
consistent with $\mathcal{R}$, i.e., the green loop configurations that
cover all the kagome edges not covered by a red loop. Thus
\begin{equation}
Z^{\rm kag}_{\rm FPL}=(a_1+a_2)^{\frac{3N}{2}} \sum_{\mathcal{R}}
x_1^{N_{x_1}}x_2^{N_{x_2}}z_1^{N_{z_1}}z_2^{N_{z_2}}(n-1)^{l_r}
\sum_{\mathcal{G}|\mathcal{R}}
y_1^{N_{y_1}}y_2^{N_{y_2}}1^{l_g} \, .
\label{partsum}
\end{equation}
For each vertex visited by green loops only, there are precisely two
possible local loop configurations.
Since the loop weight of the green loops is 1, the summation
over such pairs of configurations  is trivial:
\begin{equation}
\sum_{\mathcal{G}|\mathcal{R}}y_1^{N_{y_1}}y_2^{N_{y_2}}1^{l_g}=
\sum_{\mathcal{G}|\mathcal{R}}y_1^{N_{y_1}}y_2^{N_{y_2}}=
(y_1+y_2)^{N_g}=1 \, ,
\end{equation}
where $N_g$ is the number of green-only vertices.
The FPL partition sum thus reduces to that of a dilute loop model,
involving only red loops of weight $n-1$:
\begin{equation}
Z^{\rm kag}_{\rm FPL}(a_1,a_2,n)=
(a_1+a_2)^{\frac{3N}{2}} Z_{\rm DL}^{\rm kag}(x_1,x_2,z_1,z_2,n-1) \, ,
\label{fpldl}
\end{equation}
where the partition function of the dilute loop model is defined as
\begin{equation}
Z_{\rm DL}^{\rm kag}(x_1,x_2,z_1,z_2,n) \equiv \sum_{{\mathcal{L}}}
x_1^{N_{x_1}}x_2^{N_{x_2}} z_1^{N_{z_1}}z_2^{N_{z_2}} n^{N_l} \, ,
\label{ZDL}
\end{equation}
in which we forget the color variable, and denote the number of 
loops in a dilute configuration $\mathcal{L}$ as $N_l$.
The dilute vertices are shown in Fig.~\ref{5vsdl}, together with their
weights. The exponents of the vertex weights in Eq.~(\ref{ZDL}) represent
the numbers of the corresponding vertices.
Because of the similarity with the derivation of branch 0 on the square
lattice, we refer to the model (\ref{ZDL}) as branch 0 of the kagome
O($n$) loop model.

\begin{figure}
\includegraphics[scale=0.28]{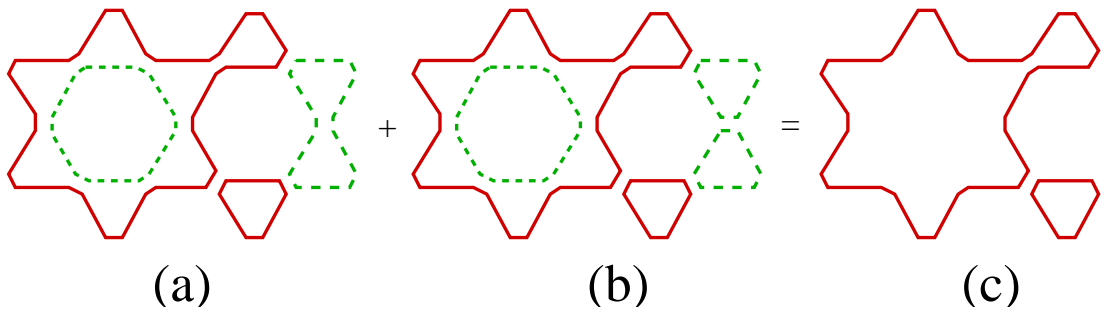}
\caption{(color online). Partial summation on the green loops.
The solid lines represent red loops, and the dashed lines green loops. 
For a fixed configuration of red loops,
each vertex visited only by green loops has two possible weights: 
$y_1$ or $y_2$ (see Fig.~\ref{6vs}).
For the simple case shown here, there are two possible configurations (a) 
and (b), of which the relative weights are $x_1^6x_2^2y_1z_1z_2(n-1)^21^2$ and 
$x_1^6x_2^2y_2z_1z_2(n-1)^21^3$ respectively. 
Addition  of these weights yields the weight $x_1^6x_2^2z_1z_2(n-1)^2$ 
of the DL configuration shown in (c).}
\label{pdl}
\end{figure}

The transformation between the FPL and the DL model is illustrated
in Fig.~\ref{pdl}.

\subsection{Dilute loop model to O($n$) spin model}
The Boltzmann weights in Eq.~(\ref{ZDL}) contain, besides the loop weights,
only local weights associated with the vertices of the kagome lattice.
Just as in the case of the O($n$) model on the square lattice described 
in Ref.~\cite{BN}, there are precisely four incoming edges at each vertex. 
This implies that there is an equivalent O($n$) spin model:
\begin{equation}
Z_{\rm DL}^{\rm kag}(x_1,x_2,z_1,z_2,n) = Z_{\rm spin}(x_1,x_2,z_1,z_2) \, ,
\end{equation}
of which the local weights have the same relation with the vertex
weights as for the square lattice model of Ref.~\cite{BN}. Thus, the
partition sum of the spin model is expressed by
\begin{displaymath}
Z_{\rm spin}(x_1,x_2,z_1,z_2) \equiv 
\int \left[ \prod_{i} d \vec{s}_i \right] \prod_{v} [1+
x_1 (\vec{s}_{v1}\cdot\vec{s}_{v2}+ \vec{s}_{v3}\cdot\vec{s}_{v4})+
\mbox{\hspace{10 mm}}
\end{displaymath}
\begin{equation}
\mbox{\hspace{20 mm}}
x_2 (\vec{s}_{v1}\cdot\vec{s}_{v4}+ \vec{s}_{v2}\cdot\vec{s}_{v3})+
z_1 (\vec{s}_{v1}\cdot\vec{s}_{v2})(\vec{s}_{v3}\cdot\vec{s}_{v4})+
z_2 (\vec{s}_{v1}\cdot\vec{s}_{v4})(\vec{s}_{v2}\cdot\vec{s}_{v3})] \, .
\label{spin}
\end{equation}
The product is on all vertices $v$ of the kagome lattice.
The spins $\vec{s}_{vi}$ sit on the midpoints of the edges of the 
kagome lattice. Their subscript ``$vi$'' specifies the vertex $v$ as
well as the position $i$ (with $1\leq i \leq 4$) with respect to the
vertex. The label $1$ runs clockwise around each vertex, such that
the spins $\vec{s}_{v1}$ and $\vec{s}_{v2}$ sit on the same side
of the honeycomb edge passing through vertex $v$. The spins have
$n$ Cartesian components and are normalized to length $\sqrt{n}$.
There are two different notations for each spin (because
each spin is adjacent to two vertices), but a given subscript $vi$
refers to only one spin. Here the number $n$ is restricted to positive
integers, of which only the case $n=1$ is expected to be critical.

\subsection{Condition for criticality}
Since the critical point of the RC model on the honeycomb lattice is
known \cite{Wufy} as a function of $q$, namely
\begin{equation}
(u_{\rm hc}^{\rm c})^3-3q(u_{\rm hc}^{\rm c})+q^2=0 \, ,
\end{equation}
the corresponding critical point of the $n=\sqrt{q}$ FPL model on 
the kagome lattice is also known. According to Eq.~(\ref{wfpl})
\begin{eqnarray}
a_1^{\rm c} &=& u_{\rm hc}^{\rm c} q^{-\frac{1}{6}} \nonumber\\
a_2^{\rm c} &=& q^{\frac{1}{3}} \, , \label{cpfpl}
\end{eqnarray}
from which the corresponding critical point of the DL model with loop
weight $n=\sqrt{q}-1$ on the kagome lattice follows as
\begin{eqnarray}
x_1^{\rm c} &=& z_1^{\rm c} =
            \frac{u_{\rm hc}^{\rm c}}{u_{\rm hc}^{\rm c}+\sqrt{q}} \nonumber\\
x_2^{\rm c} &=& z_2^{\rm c} =
            \frac{\sqrt{q}}{u_{\rm hc}^{\rm c}+\sqrt{q}} \, . \label{cpdl}
\end{eqnarray}

\section{Derivation of some critical properties}
\label{deriv}
The transformations described in Sec.~\ref{mapsec} leave (apart from
a shift by a constant) the free energy unchanged, and lead to relations
between the thermodynamic observables of the various models. Thus, the
conformal anomaly and some of the critical exponents of the FPL and the
DL models can be obtained from the existing results for the random-cluster
model. Thus, like in the analogous case of the O($n$) model on the
square lattice \cite{BN}, the FPL model on the kagome lattice should be
in the universality class of the low-temperature O($n$) phase.
However, the representation of magnetic correlations in our present 
cylindrical geometry leads to a complication.  The kagome lattice
structure, together with the FPL constraint, imposes the number of loop
segments running along the cylinder to be even.
Since the O($n$) spin-spin correlation function is represented by a
single loop segment in the loop representation, which cannot be embedded
in an FPL model on the kagome lattice, it is not clear how to represent
magnetic correlations in this model. Thus we abstain from a further
discussion of the scaling dimensions of the FPL model.

1. {\em conformal anomaly}

For the FPL model with loop weight $n$ on the kagome lattice, the 
conformal anomaly $c$ is equal to that of the $n=\sqrt q$ Potts
model \cite{BCN,Affl}:
\begin{equation}
c=1-\frac{6}{m(m+1)},~~~~ 2\cos \frac{\pi}{m+1}=n,~~~~ m \ge 1 \, .
\end{equation}
In the Coulomb gas language \cite{BN2}, it can be expressed as a function
of the Coulomb gas coupling constant $g$, with $g=m/(m+1)$:
\begin{equation}
c=1-\frac{6(1-g)^2}{g}, ~~~~2\cos(\pi g)=-n, ~~~~0 \leq g \leq 1 \, . 
\label{cfpl}
\end{equation}

The conformal anomaly $c$ of the branch-$0$ critical O($n$) DL model
on the kagome lattice with loop weight $n$ is
given by the same formula, but with $n$ replaced by $n+1$:
\begin{equation}
c=1-\frac{6}{m(m+1)}, ~~~~2\cos \frac{\pi}{m+1}=n+1, ~~~~m \ge 1 \, .
\label{cdl}
\end{equation}

The conformal anomaly is, via the number $m$, related to a set of scaling
dimensions $X_i$ as determined by the Kac formula \cite{FQS}:
\begin{equation}
X_i=\frac {[p_i (m+1)-q_i m]^2-1}{2m (m+1)} \, ,
\label{CFXi}
\end{equation}
where $p_i$ and $q_i$ are integers for unitary models. 

2. {\em temperature exponent}

For the branch-$0$ critical DL model with loop weight $n$ on the kagome
lattice, the temperature exponent is expected to be the same as
that for branch 0 on the square lattice \cite {BN}, namely $X_t=X_i$ 
with $p_i=m, q_i=m$ in Eq.~(\ref{CFXi}). 

3. {\em magnetic exponent}

The magnetic exponent of the branch-$0$ DL model with $n=0$ on the kagome
lattice is {\em not} equal to the magnetic
exponent of the low temperature O($n+1$) loop model.
The same situation was found earlier for the branch-$0$ O($0$)
model on the square lattice \cite{BN}. 
According to the reason given in \cite {BN}, the magnetic
exponent is equal to the 
temperature one, i.e., the $p_i=m,q_i=m$ entry of Eq.~(\ref{CFXi}).
The geometry of the underlying FPL model, where the number of dangling 
bonds is restricted to be even, plays here  an essential role.
Note that the magnetic exponent of the tricritical dilute O($n$)
model \cite{GNB}, even at the $\theta$ point, is different from that
of branch 0.
 
These results for $X_t$ and $X_h$ are expressed in the Coulomb gas
language as 
\begin{equation}
X_t=X_h=1-1/2g \, .
\label{xtxh}
\end{equation}

\section{Numerical verification}
\label{numver}
\subsection{ Construction of the transfer matrix} 
The transfer matrix is constructed for an $L \times M$ loop model wrapped
on a cylinder, with its axis perpendicular to one of the lattice edge
directions of the kagome lattice. The finite size $L$ is defined such that
the circumference of the cylinder is spanned by $L/2$ elementary hexagons
(corner to corner).  
The cylinder is divided into $M$ slices, of which $L$ sites form a cyclical
row, while each of the $L/2$ remaining sites forms an equilateral triangle
with two of the sites of the cyclical row.
The length of the cylinder is thus $M\sqrt{3}$. 

The partition function of this finite-size DL model is given by
Eq.~(\ref{zfpl}), but with ${\mathcal{L}}_M$ instead of ${\mathcal{L}}$,
in order to specify the length $M$ of the cylinder:
\begin{equation}
Z^{(M)}=\sum_{{\mathcal{L}}_M}x_1^{N_{x_1}}x_2^{N_{x_2}}
z_1^{N_{z_1}}z_2^{N_{z_2}} n^{N_l} \, .
\end{equation}
There are open boundaries at both ends of the cylinder,  
so that there are $L$ dangling edges connected to the vertices on
row $1$, as well as on row $M$.
The way in which the end points of the dangling edges are pairwise
connected by the loop configuration ${\mathcal{L}}_M$ is defined as the
`connectivity', see Ref.~\cite{BN} for details.
Here we ignore the dangling edges of row 1 (except for a topological
property that will be considered later) and focus on the $L$ dangling
edges of row $M$.
Since it is determined by the loop configuration, the connectivity
$\beta$ at row $M$ is written
as a function of ${\mathcal{L}}_M$: $\beta=\varphi({\mathcal{L}}_M)$.
The partition sum is divided into a number of restricted sums
$Z_{\beta}^{(M)}$, each of which collects all terms in $Z^{(M)}$ having 
connectivity $\beta$ on row $M$, i.e.:
\begin{equation}
Z^{(M)}=\sum_{\beta}Z^{(M)}_{\beta} \, , \mbox{\hspace{10mm}}
Z^{(M)}_{\beta}=\sum_{{\mathcal{L}}_M}\delta_{\beta,\varphi({\mathcal{L}}_M)}
x_1^{N_{x_1}}x_2^{N_{x_2}}z_1^{N_{z_1}}z_2^{N_{z_2}}n^{N_l} \, .
\label{psum}
\end{equation}
An increase of the system length $M$ to $M+1$ leads to a new
configuration ${\mathcal{L}}_{M+1}$ which can be decomposed in
${\mathcal{L}}_M$ and the appended configuration $l_{M+1}$ on row $M+1$.
The graph $l_{M+1}$ fits the dangling edges of the loop graph
${\mathcal{L}}_M$ on the $M$-row lattice. The addition of the new row
increases the number of the four kinds of vertices and of the number
of loops by $n_{x_1}$, $n_{x_2}$, $n_{z_1}$, $n_{z_2}$ and $n_l$
respectively.
The restricted partition sum of the system with $M+1$ rows is
\begin{displaymath}
Z_{\alpha}^{(M+1)}=
\sum_{{\mathcal{L}}_{M+1}}\delta_{\alpha,\varphi({\mathcal{L}}_{M+1})}
x_1^{N_{x_1}+n_{x_1}}x_2^{N_{x_2}+n_{x_2}}z_1^{N_{z_1}+n_{z_1}}
z_2^{N_{z_2}+n_{z_2}}n^{N_l+n_l}=
\end{displaymath}
\begin{equation}
\sum_{{\mathcal{L}}_M}x_1^{N_{x_1}}x_2^{N_{x_2}}
z_1^{N_{z_1}}z_2 ^{N_{z_2}}n^{N_l}\\
\sum_{{l}_{M+1}|{\mathcal{L}}_M}\delta_{\alpha,\varphi({\mathcal{L}}_{M+1})}
x_1^{n_{x_1}}x_2^{n_{x_2}}z_1^{n_{z_1}}z_2^{n_{z_2}}n^{n_l} \, .
\label{zm1}
\end{equation}
The last sum is on all sub-graphs ${l}_{M+1}$ that fit ${\mathcal{L}}_M$. 
The connectivity $\varphi({\mathcal{L}}_{M+1})$ depends only on the 
connectivity $\beta$ on row $M$, and on $l_{M+1}$, 
so that we may write $\varphi({\mathcal{L}}_{M+1})=\psi(\beta,l_{M+1})$.
Thus Eq.~(\ref{zm1}) assumes the form
\begin{equation}
Z_{\alpha}^{(M+1)}=\sum_{\beta}\sum_{{\mathcal{L}}_M}
\delta_{\beta,\varphi({\mathcal{L}}_M)}x_1^{N_{x_1}}x_2^{N_{x_2}}
z_1^{N_{z_1}}z_2^{N_{z_2}}n^{N_l}\\
\sum_{l_{M+1}|\beta}\delta_{\alpha,\psi(\beta,l_{M+1})}
x_1^{n_{x_1}}x_2^{n_{x_2}}z_1^{n_{z_1}}z_2^{n_{z_2}} n^{n_l} \, .
\label{lrec}
\end{equation}
The third sum depends only on $\alpha$ and $\beta$, and thus defines
the elements of the transfer matrix $\mathbf T$ as 
\begin{equation}
T_{\alpha\beta} \equiv \sum_{l_{M+1}|\beta}\delta_{\alpha,\psi(\beta,l_{M+1})}
x_1^{n_{x_1}}x_2^{n_{x_2}}z_1^{n_{z_1}}z_2^{n_{z_2}} n^{n_l} \, ,
\end{equation}
Substitution of $T_{\alpha\beta}$ and Eq.~(\ref{psum}) in Eq.~(\ref{lrec})
then yields the recursion of the restricted partition sum as
\begin{equation}
Z_{\alpha}^{(M+1)}=\sum_{\beta}T_{\alpha\beta}Z_{\beta}^{(M)} \, .
\end{equation}

\begin{figure}
\includegraphics[scale=0.35]{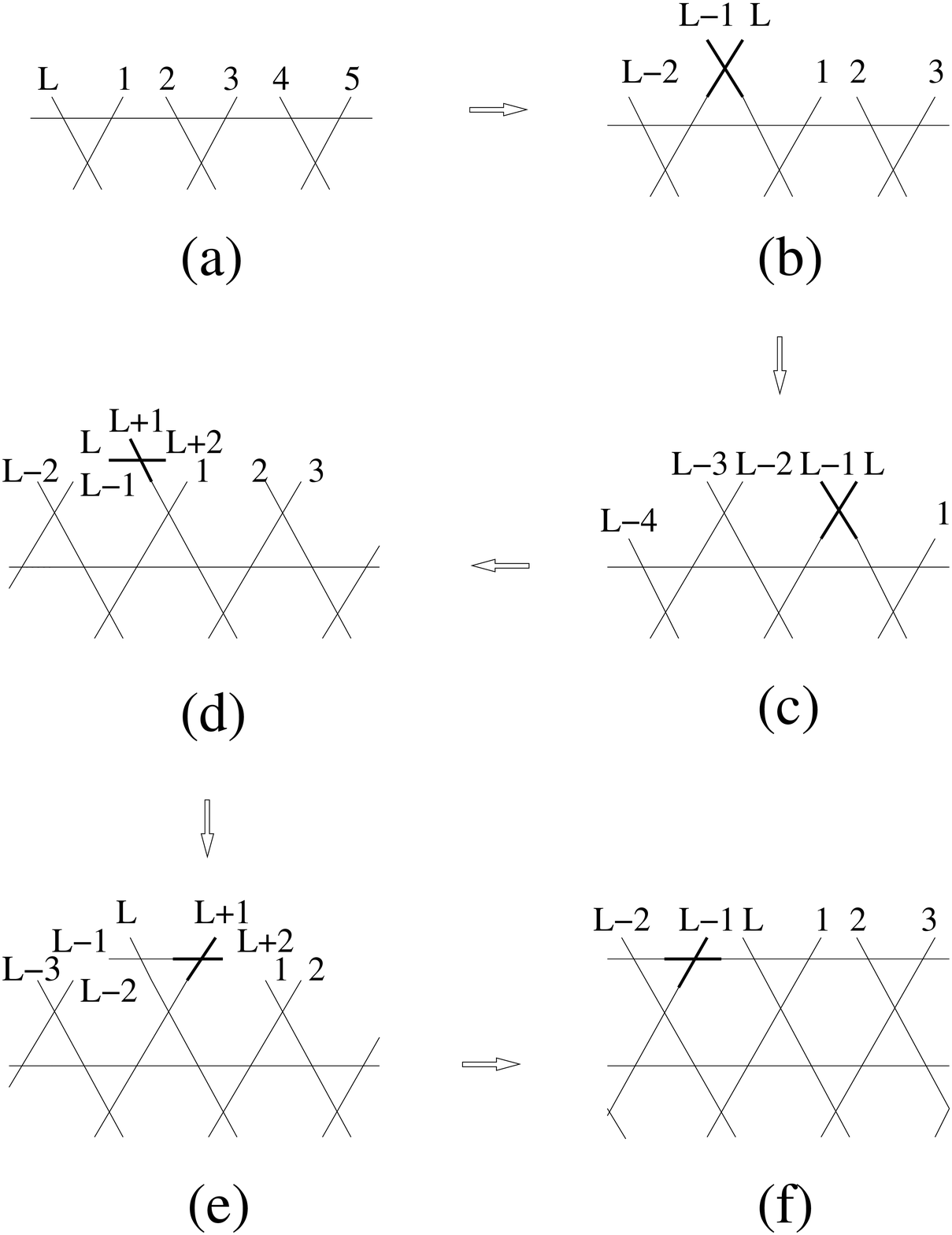}
\caption{Constructing the transfer matrix. 
Appending a new row to the configuration is achieved in two parts. 
The first part consists of $L/2$ steps and is denoted $T_{L/2}\ldots T_1$
(which are executed from right to left). Each step adds a new site to the
lattice. Two of these steps are illustrated in (a) to (c).
The number of dangling bonds does not change during these steps.  The
second part consists of $L$ steps and is denoted $T_{3L/2}\ldots T_{L/2+1}$.
The first step of these, $T_{L/2+1}$, adds a new vertex to the sub-row
and increases the number of dangling bonds by 2 as shown in (d). The
following steps $T_{L/2+2} \cdots T_{3L/2-1}$ append vertices sequentially,
and do not change 
the number of dangling bonds. After adding the last vertex by $T_{3L/2}$ to 
the sub-row, the construction of a new row has been completed and the
size of the system shrinks from $L+2$ to $L$.
}
\label{TM}
\end{figure}

In order to save memory and computer time, the transfer matrix of a system
with finite size $L$ is decomposed in $\frac{3L}{2}$ sparse matrices:
\begin{equation}
T=T_{\frac{L}{2}+L}\cdot T_{\frac{L}{2}+L-1}\cdot \ldots \cdot 
T_{\frac{L}{2}+1} \cdot T_{\frac{L}{2}}\cdot T_{\frac{L}{2}-1} \cdot
\ldots \cdot T_2 \cdot T_1 \, , 
\end{equation}
where $T_i$ denotes an operation which adds a new vertex $i$ on a new row, 
as illustrated in Fig.~\ref{TM}. Most of these sparse matrices are square,
but $T_{\frac{L}{2}+1}$ is not, because it increases the number of dangling
bonds by two. The action of the other rectangular matrix, $T_{\frac{L}{2}+L}$,
reduces the number of dangling bonds again to $L$.

During the actual calculations, we only store the positions and values of
the non-zero elements of a sparse matrix, in a few one-dimensional arrays.
Moreover, this need not be done for all the sparse  matrices, because
there are only four independent matrices. The other ones are related to
these by the action of the translation operator \cite{BNFSS,BN}.

While the construction of the transfer matrix is formulated in terms of
connectivities on the topmost rows $M$ and $M+1$, the connectivity on
row 1 is not entirely negligible. In particular, the number of dangling
loop segments on that row can be even or odd. As a consequence the
number of dangling loop segments on the topmost row is then also even or
odd respectively. This leads to a decomposition of the transfer matrix
in an even and an odd sector. The odd sector corresponds with a single
loop segment running in the length direction of the cylinder.

\subsection{Results of the numerical calculation}

For a model on an infinitely long cylinder with finite size $L$, the free 
energy per unit of area is determined by 
\begin{equation}
f(L)=\frac{1}{\sqrt{3}L}\ln \Lambda _L^{(0)} \, ,
\end{equation}
where $\Lambda _L^{(0)}$ is the largest eigenvalue of $T$ in the
$n_d=0$ sector. From the finite-size data for $f(L)$ we estimated the
conformal anomaly $c$ \cite{BCN}.

The magnetic correlation length $\xi_h(L)$ is related to the magnetic
gap in the eigenvalue spectrum of $T$ as
\begin{equation}
\xi _h^{-1}(L)=\frac{1}{\sqrt{3}}\ln (\Lambda_L^{(0)}/\Lambda_L^{(1)}) \, ,
\end{equation}
where $\Lambda_L^{(1)}$ is the largest eigenvalue of $T$ in the $n_d=1$
sector.

The temperature correlation length $\xi_t(L)$ is related to the
temperature gap in the eigenvalue spectrum of $T$ as
\begin{equation}
\xi _t^{-1}(L)=\frac{1}{\sqrt{3}}\ln (\Lambda_L^{(0)}/\Lambda_L^{(2)}) \, ,
\end{equation}
where $\Lambda_L^{(2)}$ is the second largest eigenvalue of $T$ in the 
$n_d=0$ sector.
Using Cardy's conformal mapping \cite{Cardy-xi} of an infinite cylinder on 
the infinite plane, one can thus estimate the temperature dimension 
$X_t$ and $X_h$. 

We calculated the finite-size data for the free energies of the FPL 
model at the critical
points given by Eq.~(\ref{cpfpl}) for system sizes $L=2,~4,~\cdots,~28$.
These data include the case $n=0$; this is possible because, for
$q \to 0$ one has $u_{\rm hc}^{\rm c}=\sqrt{3q}$, so that the
ratio between $a_1^{\rm c}$ and $a_2^{\rm c}$ in Eq.~(\ref{cpfpl})
remains well defined in this limit. 

The additional loop configurations allowed by the dilute model lead to
a larger transfer matrix for a given system size, so that our results
at the critical points given by Eqs.~(\ref{cpdl}) are restricted to
sizes $L=2,~4,~\cdots,~18$. The latter results also include the
temperature and magnetic gaps. 

The finite-size data for the FPL and DL models displayed a good apparent
convergence, and were fitted using
methods explained earlier \cite{BNFSS,BN,GNB}, see also Ref.~\cite{FSS}.

In the kagome lattice FPL model, it is not possible to introduce one
single open loop segment running in the length direction of the cylinder.
The presence of a single chain would force unoccupied edges into the
system, in violation of the FPL condition. Therefore, we have no results
for $X_h$. Furthermore, in the case of the low-temperature O($n$) phase,
the eigenvalue associated with $X_t$ decreases rapidly when $n$ becomes 
smaller than 2, and becomes dominated by other eigenvalues.
Therefore, also results for $X_t$ are absent for the FPL model, and
our results are here restricted to the  conformal anomaly $c$. 
The resulting estimates for the FPL model are listed in Tab.~\ref{tab1}.

\begin{table}
\caption{Conformal anomaly $c$ of the FPL model as determined by
the transfer-matrix calculations described in the text. 
The sizes of the system $L$ are from $2$ to $28$. Estimated error
margins in the last decimal place are given in parentheses. 
The numerical results are indicated by `num'. For comparison, we include
theoretical values indicated by `th', as given by Eq.~(\ref{cfpl}).
 }
\label{tab1}
\begin{center}
\begin{tabular}{||l|l|l||}
\hline
$n$        & $c^{\rm th}$  & $c^{\rm num}$    \\
\hline
$0   $     & $-2$          & $-2.000001$   (1)\\
$0.25$     & $-1.3526699$  & $-1.352670$   (5)\\
$0.5 $     & $-0.8197365$  & $-0.819737$   (5)\\
$0.75$     & $-0.3749081$  & $-0.374908$   (5)\\
$1   $     & $0         $  & $ 0$             \\
$1.25$     & $0.31782377$  & $ 0.31782$    (2)\\
$\sqrt{2}$ & $1/2       $  & $ 0.5000000$  (2)\\
$1.50$     & $0.58757194$  & $ 0.587565$   (5)\\
$\sqrt{3}$ & $4/5       $  & $ 0.80000$    (1)\\
$1.75$     & $0.81497930$  & $ 0.81498$    (2)\\
$2   $     & $1         $  & $ 1.0001$     (1)\\
\hline
\end{tabular}
\end{center}
\end{table}

The results for the eigenvalue $\Lambda_L^{(0)}$ of the the FPL model
satisfy, within the numerical precision in the order of $10^{-12}$,
the relation between the  FPL and DL models derived in Sec.~\ref{fptodl}.
The larger dimensionality of the transfer matrix of the DL model in
comparison with the FPL model generates new eigenvalues, and thus leads
to new scaling dimensions that are absent in the FPL model.
Final estimates for the conformal anomaly $c$ and for the scaling
dimensions $X_t$ and $X_h$ are listed in Tab.~\ref{tab2} for the DL model.
They agree well with the theoretical predictions, which are included in
the table.  Here we recall that, in analogy with the case of the
branch-$0$ O($n$) loop model on the square lattice \cite{BN}, the 
magnetic scaling dimension should be exactly equal to the thermal one.
This is in agreement with our numerical results. We found that
the eigenvalues $\Lambda_L^{(1)}$ and $\Lambda_L^{(2)}$ were the same
within the numerical error margin. Thus, we list only one column with
results for the exponents in Tab.~\ref{tab2}.

\begin{table}
\caption{Conformal anomaly $c$, magnetic scaling dimension $X_h$ and
temperature scaling dimension $X_t$ of 
the DL model as determined by the transfer-matrix calculations described 
in the text.  Estimated error
margins in the last decimal place are given in parentheses. The numerical
results are indicated by `num'. For comparison, we include the theoretical
values indicated by `th', as given by Eqs.~(\ref{cdl}) and (\ref{xtxh}).
 }
\label{tab2}
\begin{tabular}{||l|l|l|l|l||}
\hline
$n$ &$c^{\rm th}$ &$c^{\rm num}$&$X_h^{\rm th}$, $X_t^{\rm th}$ &
$X_h^{\rm num}$, $X_t^{\rm num}$\\
\hline
$-1   $     &$-2        $&$-2.0000$  (5) &$0$          &$0.0000000$ (1) \\
$-0.75$     &$-1.3526699$&$-1.3524$  (3) &$0.073890718$&$0.0738908$ (2) \\
$-0.5$      &$-0.8197365$&$-0.8194$  (5) &$0.138570601$&$0.138571$  (1) \\
$-0.25$     &$-0.3749081$&$-0.3747$  (3) &$0.196602972$&$0.196605$  (5) \\
$0    $     &$0         $&$ 0$           &$1/4        $&$0.25000$   (1) \\
$0.25$      &$0.31782377$&$ 0.31778$ (5) &$0.300602502$&$0.30061$   (5) \\
$\sqrt{2}-1$&$1/2       $&$ 0.500001$ (1)&$1/3        $&$0.33334$   (1) \\
$0.50$      &$0.58757194$&$ 0.5876 $ (1) &$0.350604267$&$0.35061$   (1) \\
$\sqrt{3}-1$&$4/5       $&$ 0.8002 $ (3) &$2/5        $&$0.3997 $   (5) \\
$0.75$      &$0.81497930$&$ 0.8151 $ (3) &$0.404150985$&$0.4037 $   (5) \\
$1 $        &$1         $&$ 1.002  $ (3) &$1/2        $&$0.48$      (3) \\
\hline
\end{tabular}

\end{table}

\section{conclusion}
\label{concl}
We found a branch of critical points of the dilute loop model on the
kagome lattice as a function of the loop weight $n$, which is related to
the $q=(n+1)^2$-state Potts model on the honeycomb lattice. The critical
properties of these critical points are conjectured and verified by
numerical transfer matrix calculations and a finite-size-scaling analysis.
As expected, the model falls into the same universality class as branch
$0$ of the O($n$) loop model \cite{BN} on the square lattice.
The analysis did, however, yield a difference. This is due to the geometry
of the lattice. For the square lattice, it was found \cite{BN} that there
exists a magnetic scaling dimension $X_{\rm int,1}$ as revealed by the
free-energy difference between even and odd systems. Such an alternation
is absent in the free-energy of the present model on the kagome lattice.
While the number of dangling edges may be odd or even for the square
lattice, it can only be even in the present case of the kagome lattice.

The numerical accuracy of the results for the conformal anomaly and
the exponents is much better than what can be typically achieved 
for an arbitrary critical point, whose location in the parameter 
space has to be determined in advance by so-called phenomenological
renormalization \cite{MPN}. This seems not only due to the limited
precision of such a critical point. We suppose that the main reason is
that irrelevant scaling fields tend to be suppressed in exactly 
solvable parameter subspaces.

\acknowledgments
We are much indebted to Bernard Nienhuis, for making his insight
in the physics of O($n$) loop models available to us.
This research is supported by the National Science Foundation of
China  under Grant \#10675021, by the Beijing Normal University
through a grant as well as support from its HSCC (High Performance
Scientific Computing Center), and by the Lorentz Fund (The
Netherlands).

\end{document}